\documentstyle[prb,aps,multicol,epsf]{revtex}

\ifpreprintsty\relax\else
\newlength{\colwidth}
\fi

\begin{document}
\draft 
\title{Thermal Crossover between Ultrasmall Double and Single Junction}
\author{Heinz--Olaf M{\"u}ller\cite{email}} 
\address{Institut f{\"u}r Festk{\"o}rperphysik,
  Friedrich--Schiller--Universit{\"a}t, 
  Lessingstra{\ss}e 8, DE--07743 Jena, Germany} 
\date{\today}

\maketitle
\begin{abstract}
The crossover from double-junction behavior to single-junction
behavior of ultrasmall tunnel junctions is studied theoretically in a
scanning-tunneling microscope setup. The independently variable tip
temperature of the microscope is used to monitor the transition
between both regimes. 
\end{abstract}

\ifpreprintsty\relax\else\begin{multicols}{2}\narrowtext\fi

\section{Introduction}

The Coulomb blockade of the current through a system of ultrasmall
tunnel junctions is one of the basic features of single
electronics. There are two simple systems capable of producing this
blockade if connected to a voltage source, namely the single junction
within a high-impedant environment and the double junction. Owing to
their different features the theoretical description of the former is
much more simple than that of the latter, whereas manufacturing
favors the latter instead of the former.

In this paper we study theoretically the crossover from
double-junction behavior to single-junction behavior in the usually
scanning-tunneling-microscope setup.\cite{wil1} The tip junction,
i.e.\ where the outer electrode is formed by the microscope tip, is
characterized by the resistance $R_1$ and the capacitance $C_1$,
whereas the corresponding parameters of the other junction are denoted
$R_2$ and $C_2$. The crossover between double-junction regime and
single-junction regime is governed by the temperature of the
microscope tip, $T_{\rm tip}$, which is thought to be independent of
the temperature of the other electrodes, $T_{\rm base}$. We discuss
the case $T_{\rm tip}\ge T_{\rm base}$ only. In order to observe a
Coulomb blockade the base temperature is restricted to values $T_{\rm
base}\ll E_{\rm C}/k_{\rm B}$, where $E_{\rm C}=1/2\,e^2/(C_1+C_2)$ is
the Coulomb energy of the double junction.

If the tip temperature is comparable to the base temperature, a usual
double-junction behavior is expected. Its theoretical description is
the so-called ``orthodox theory'':\cite{ave3,amm2} a master equation
describes the occupation probability of different charge states of the
small island between the two junctions. The current through the system
is expressed in terms of these probabilities.

If the tip temperature is well above $T_{\rm base}$, thermal
fluctuations will make the tip junction more transparent. However, its
resistance is still large on the scale of the quantum resistance
$R_{\rm k}={\rm h}/e^2\approx25.8{\rm k}\Omega$. Thus we have
the typical single-junction setup: an ultrasmall tunnel junction at
base temperature and an high impedant almost ohmic environment
resistance. The small capacitance of the latter is negligible. 

This paper is organized as follows. The theoretical approximations for
our discussion are given in the next Section. After this follow the
discussion itself and the conclusion.

\section{Theoretical description}

\subsection{Double junction}

The ultrasmall double junction is described in terms of standard
``orthodox theory''.\cite{ave3,amm2} Assuming the tunnel rates through
the first ($1$) and second ($2$) junction in right ($r$) and left
($l$) direction are known as $r_{1,2}(n)$ and $l_{1,2}(n)$ for $n$
extra charges on the central island, the time dependence of the
occupation probability $\sigma(n)$ of the state $n$ follows from the
master equation\cite{ave3,amm2}
\begin{eqnarray}
\label{me}
\frac{{\rm d}\sigma(n,t)}{{\rm d}t} & = &
\big[r_1(n-1)+l_2(n-1)\big]\sigma(n-1,t)\\
& & + \big[r_2(n+1)+l_1(n+1)\big]\sigma(n+1,t)\nonumber\\
& & -\big[r_1(n)+r_2(n)+l_1(n)+l_2(n)\big]\sigma(n,t).\nonumber
\end{eqnarray}
For the considered stationary situation we use the stationary solution
of this equation\cite{amm2,seu1}
\begin{eqnarray}
\label{sme}
\sigma(n) & = & \frac{1}{\cal
Z}\prod_{i=-\infty}^{n-1}\big[r_1(i)+l_2(i)\big]\\
& & \times\prod_{j=n+1}^{\infty}\big[r_2(j)+l_1(j)\big]\nonumber
\end{eqnarray}
with an appropriate normalization ${\cal Z}$ so that
\[
\sum_{n=-\infty}^{\infty}\sigma(n) = 1.
\]
The stationary average current results from $\sigma(n)$
\begin{eqnarray}
\label{cme}
\langle I\rangle & = & e\sum_{n=-\infty}^{\infty}
\big[r_1(n)-l_1(n)\big]\sigma(n)\\
& = & e\sum_{n=-\infty}^{\infty}\big[r_2(n)-l_2(n)\big]
\sigma(n).\nonumber
\end{eqnarray}

Let us consider the transition rates $r_{1,2}(n)$ and $l_{1,2}(n)$ in
detail now. They depend on $n$ via the energy differences
$E_{1,2}^{r,l}(n)$ caused by the respective tunneling event,
\begin{eqnarray}
E_{1,2}^r(n) & = & E_{\rm C}\big(\frac{C_{2,1}V}{e}\mp n
-\frac{1}{2}\big)\\
E_{1,2}^l(n) & = & E_{\rm C}\big(-\frac{C_{2,1}V}{e}\pm n
-\frac{1}{2}\big).
\nonumber
\end{eqnarray}
If the electron temperature on both sides of the junction is the same,
Fermi's Golden Rule results in\cite{ave3,amm2}
\begin{equation}
\label{rateeq}
\{r,l\}_{1,2}(n) = \frac{1}{e^2R_{1,2}}\,
\frac{E_{1,2}^{r,l}(n)}{1-\exp[-\beta_{\rm base} E_{1,2}^{r,l}(n)]},
\end{equation}
where $\beta_{\rm base}=(k_{\rm B}T_{\rm base})^{-1}$ is used. In case
of considerably different temperature on both sides of the junction we
use instead\cite{kau1}
\begin{equation}
\label{ratene}
\{r,l\}_{1,2}(n) = \frac{k_{\rm B}T_{\rm tip}}{e^2 R_{1,2}}\,
\log\left[\exp\left(\frac{E_{1,2}^{r,l}(n)}{k_{\rm B}T_{\rm tip}}\right)
+1\right].
\end{equation}
As shown in Ref.~\onlinecite{kau1}, (\ref{ratene}) is a reasonable
approximation for $T_{\rm tip}\ge2T_{\rm base}$. Eq.~\ref{ratene}
indicates that the tunneling behavior in this case is governed by the
higher temperature. The influence of the temperature difference
results in the change of (\ref{ratene}) in comparison to
(\ref{rateeq}). 

\subsection{Single junction}

Since the single-junction regime requires a warm tip we consider the
hot environment only. In this case the energy excitation probability
of the environment $P(E)$ is given by a Gaussian\cite{ing2}
\begin{equation}
\label{pehot}
P(E) = \frac{1}{2}\,
\sqrt{\frac{\beta_{\rm env}}{\pi E_{\rm c}}}\,
\exp\left[-\frac{\beta_{\rm env}}{4 E_{\rm c}}\,(E-E_{\rm c})^2\right]
\end{equation}
where we introduced the single-junction charging energy $E_{\rm c} =
e^2/(2C_2)$ and $\beta_{\rm env}=(k_{\rm B}T_{\rm env})^{-1}$
describes the elevated environment temperature. Eq.~\ref{pehot}
fulfills the sum rules\cite{ing1} 
\begin{eqnarray}
\label{sumrules}
\int{\rm d}E\, P(E) & = & 1\\
\int{\rm d}E\, E\, P(E) & = & E_{\rm c}.\nonumber
\end{eqnarray}

In Ref.~\onlinecite{ing4} a Lorentzian shape is derived for the hot
environment and negligible charging energy, but we have not found
satisfactory results with that formula. In the considered temperature
range charging effects are still essential as can be seen from the
occurance of the Coulomb blockade. Hence, the situation here might be
well outside the scope of the Lorentzian formula of
Ref.~\onlinecite{ing4}. 

The single junction at $T_{\rm base}$ is described by two rates, namely
$r_2$ and $l_2$. Their functional dependence on the energy difference
between both sides of the junction follows (\ref{rateeq}), but the
energies $E_2^{r,l}$ are the voltage drops across the junction.
For low base temperature we can make use of the
approximation\cite{kre11} 
\[
\frac{E}{1-\exp(-\beta E)} \approx E\Theta(E)
+\frac{1}{\beta}\,\exp(-\gamma|E|)
\]
with $\gamma=(6/\pi^2)\beta=0.607\,927\beta$ from the normalization
\[
\int\limits_{-\infty}^0\frac{{\rm d}E\,E}{1-\exp(-\beta E)}
= \frac{\pi^2}{6\,\beta^2}
= \int\limits_{-\infty}^0\frac{{\rm d}E}{\beta}\exp(-\gamma|E|).
\]

This allows for an analytic expression of the current through the
junction
\[
I(V_2) = e\int{\rm d}E\, P(E)\big[r_2(eV_2-E)-l_2(-eV_2-E)\big],
\]
where we use $E_2^{r,l}=\pm eV_2-E$. The current is calculated in
terms of the voltage $V_2$ across the single junction, which does not
include the voltage drop $V_1$ at the tip junction. Using $g=R_{\rm
k}/R_1$ the final expression is 

\ifpreprintsty\relax\else\end{multicols}\widetext\rule{\colwidth}{0.4pt}
\hfill\fi

\begin{eqnarray}
\label{ihot}
I(V_2) & = & \frac{V_2}{2R_2}+\frac{1}{e\,R_2}\Big\{
-\frac{E_{\rm c}}{\pi}\arctan\frac{\beta_{\rm env}g}{2\pi}
(eV_2-E_{\rm c})
 +\frac{eV_2-E_{\rm c}}{\pi}\arctan\frac{\beta_{\rm env}g}{2\pi}
(eV_2+E_{\rm c})\\
& & +\frac{1}{\beta_{\rm env}g}\log
\frac{1+\big[\beta_{\rm env}g(eV+E_{\rm c})/(2\pi)\big]^2}
{1+\big[\beta_{\rm env}g(eV-E_{\rm c})/(2\pi)\big]^2}\nonumber\\
& & +\frac{4}{\beta_{\rm base}\beta_{\rm env}\gamma_{\rm base}g}
\Big[\frac{1}{[2\pi/(\beta_{\rm env}g)]^2+[eV_2-E_{\rm c}]^2}-
\frac{1}{[2\pi/(\beta_{\rm env}g)]^2+[eV_2+E_{\rm c}]^2}\Big]\Big\}.
\nonumber
\end{eqnarray}

\ifpreprintsty\relax\else\hfill
\rule{\colwidth}{0.4pt}\begin{multicols}{2}\narrowtext\fi

For dominating bias, $eV_2\gg E_{\rm c},k_{\rm B}T_{\rm env}$, the
asymptotic behavior $I(V_2)=(V_2-E_{\rm c}/e)/R_2$ is recovered.

\section{Discussion}

\ifpreprintsty\relax\else
\ifx\epsfxsize\undefined\relax\else
\begin{figure}
\epsfxsize=\columnwidth
\epsffile{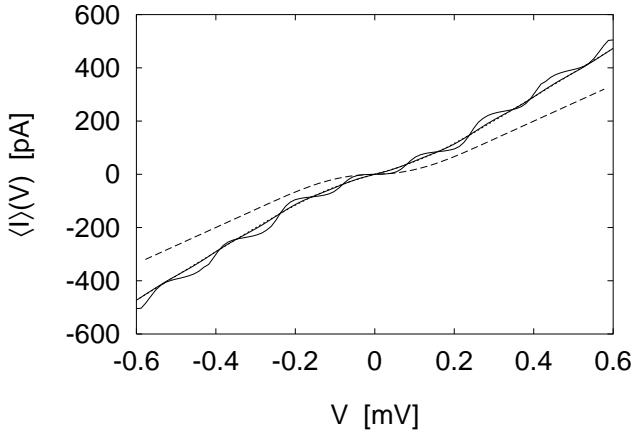}
\caption{Calculated current through an STM double junction ($C_1=1.0{\rm
fF}$, $C_2=0.1{\rm fF}$, $R_1=1.0{\rm M}\Omega$, $R_2=0.1{\rm M}\Omega$,
$T_{\rm base}=0.1{\rm K}$) for tip temperature $T_{\rm tip}=0.2{\rm
K}$ (solid line, staircase) and $T_{\rm tip}=1.0{\rm K}$ (solid line,
straight). The long dashed line shows the result of a
zero-temperature-environment calculation, whereas the short-dashed
line displays Eq.~\ref{ihot} with $T_{\rm env}=0.75{\rm K}$.}
\label{fig1}
\end{figure}
\fi
\fi

In Fig.~\ref{fig1} the result of our calculation are shown for the
case of an asymmetric double junction. Double junction systems where
an STM forms one junction are very often asymmetric. This asymmetry
results in a Coulomb staircase as seen for the low temperature curve
in Fig.~\ref{fig1}. For higher tip temperature the stairs are smeared
out, but the Coulomb blockade survives. This is the expected behavior
of a single junction in a high-impedant environment. The curve is
fitted well by our model of the hot environment and the derived
current (\ref{ihot}). The fit parameter is the temperature $T_{\rm
env}$, which is found between the low temperature $T_{\rm base}$ and
the hot $T_{\rm tip}$. Thus, it does not take wonder that the
displayed zero-temperature approximation of $P(E)$ cannot describe the
the double-junction system.

\ifpreprintsty\relax\else
\ifx\epsfxsize\undefined\relax\else
\begin{figure}
\epsfxsize=\columnwidth
\epsffile{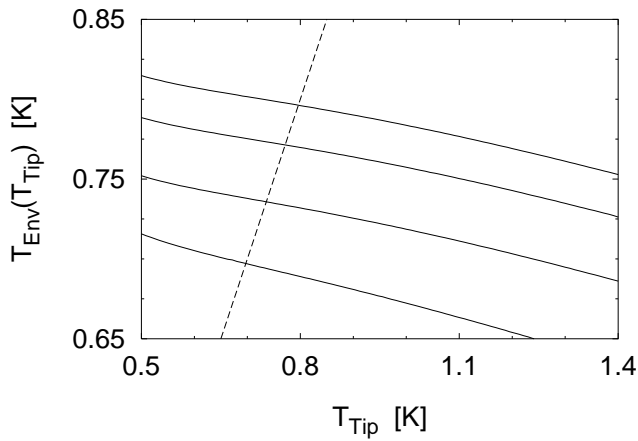}
\caption{The environment temperature $T_{\rm env}$ in dependence on
the tip temperature $T_{\rm tip}$ for four different base
temperatures $T_{\rm base}=0.05{\rm K}$, $0.1{\rm K}$, $0.15{\rm K}$,
$0.2{\rm K}$ (from the top) as resulted from a least-square fit. The
dashed line displays $T_{\rm env}=T_{\rm tip}$.}
\label{fig2}
\end{figure}
\fi
\fi

An interesting remaining question is the dependence of the environment
temperature $T_{\rm env}$ on the choice of the base temperature
$T_{\rm base}$ and the tip temperature $T_{\rm tip}$. Since $T_{\rm
env}$ is a theoretical parameter of our model we investigated this
dependence by means of a fit procedure. The results (for the system of
Fig.~\ref{fig1}) are shown in Fig.~\ref{fig2}. $T_{\rm env}$ is well
above $T_{\rm base}$, which corresponds to the results found in
Ref.~\onlinecite{kau1} for the single-electron electrometer. We
restrict the consideration to the case $T_{\rm env}\le T_{\rm tip}$. 
For low tip temperatures the Coulomb staircase makes a fit meaningless.
The dependence on $T_{\rm tip}$ is weak. Nevertheless, it is
astonishing that $T_{\rm env}$ decreases with increasing $T_{\rm
tip}$ and increasing $T_{\rm base}$. An rigorous treatment of the
dependence of the (theoretical) environment temperature on the (real)
base and tip temperature requires further investigation.

\section{Conclusion}

In a theoretical treatment we have shown that the crossover between
double-junction behavior and single-junction behavior in an STM setup
can be monitored by a single parameter, the tip temperature of the
STM. The tip junction works as an ohmic environment resistance for the
hot tip. Even if the temperature on both sides of this junction is
very different, its behavior can be described by an environment
temperature alone. 

\section*{Acknowledgment}

Extensive help with the fit procedure provided by Martin Springer of
Lund University, Sweden, is gratefully acknowledged.


\ifpreprintsty\narrowtext\else\relax\fi

\ifx\epsfxsize\undefined
\begin{figure}
\caption{Calculated current through an STM double junction ($C_1=1.0{\rm
fF}$, $C_2=0.1{\rm fF}$, $R_1=1.0{\rm M}\Omega$, $R_2=0.1{\rm M}\Omega$,
$T_{\rm base}=0.1{\rm K}$) for tip temperature $T_{\rm tip}=0.2{\rm
K}$ (solid line, staircase) and $T_{\rm tip}=1.0{\rm K}$ (solid line,
straight). The long dashed line shows the result of a
zero-temperature-environment calculation, whereas the short-dashed
line displays Eq.~\ref{ihot} with $T_{\rm env}=0.75{\rm K}$.}
\label{fig1}
\end{figure}
\else\relax\fi

\ifx\epsfxsize\undefined
\begin{figure}
\caption{The environment temperature $T_{\rm env}$ in dependence on
the tip temperature $T_{\rm tip}$ for four different base
temperatures $T_{\rm base}=0.05{\rm K}$, $0.1{\rm K}$, $0.15{\rm K}$,
$0.2{\rm K}$ (from the top) as resulted from a least-square fit. The
dashed line displays $T_{\rm env}=T_{\rm tip}$.}
\label{fig2}
\end{figure}
\else\relax\fi

\ifpreprintsty\relax\else\end{multicols}\widetext\fi

\end{document}